\documentclass[aps,prl,amsmath,amssymb,reprint,twocolumn,superscriptaddress,showpacs]{revtex4-1}
\usepackage{epsfig,amssymb}
\usepackage{graphicx}
\usepackage{dcolumn}
\usepackage{bm}
\usepackage{dcolumn}
\usepackage{multirow}
\begin{document}
\title{Dispersion and lineshape of plasmon satellites in
  one, two and three dimensions}

\author{Derek~Vigil-Fowler}
\affiliation{National Renewable Energy Laboratory, Golden, Colorado
  80401, USA.}
\author{Steven G. Louie}
\affiliation{Department of Physics, University of California at
  Berkeley and Materials Science Division, Lawrence Berkeley National
  Laboratory, Berkeley, CA 94720, USA.}
\author{Johannes~Lischner}
\email{jlischner597@gmail.com}
\affiliation{Department of Physics and Department of Materials, and
  the Thomas Young Centre for Theory and Simulation of Materials,
  Imperial College London, London SW7 2AZ, United Kingdom}


\begin{abstract}
  Using state-of-the-art many-body Green's function calculations based
  on the ``GW plus cumulant'' approach, we analyze the properties of
  plasmon satellites in the electron spectral function resulting from
  electron-plasmon interactions in one-, two- and three-dimensional
  systems. Specifically, we show how their dispersion relation,
  lineshape and linewidth are related to the properties of the
  constituent electrons and plasmons. To gain insight into the
  many-body processes giving rise to the formation of plasmon
  satellites, we connect the ``GW plus cumulant'' approach to a
  many-body wavefunction picture of electron-plasmon interactions and
  introduce the coupling-strength weighted electron-plasmon
  joint-density states as a powerful concept for understanding plasmon
  satellites.
\end{abstract}

\maketitle

\emph{Introduction}.---The interaction of electrons with bosons is of
fundamental importance for many phenomena in condensed matter physics,
plasma physics and cold atom physics. Recently, there has been great
interest in the coupling of electrons and plasmons, which are
collective excitations describing quantized oscillations of the charge
density. For example, the decay of plasmons into energetic or ``hot''
electron-hole pairs in metallic surfaces and nanoparticles, which is
triggered by electron-plasmon coupling, has led to a new generation of
plasmonic devices for photovoltaics and photocatalysis
\cite{mukherjee2012hot,clavero2014plasmon,moskovits2015case}.

Satellite features in the spectral function of electrons are another
consequence of electron-plasmon interactions. Such plasmon satellites
have long been known in core-electron photoemission spectra
\cite{lundqvist1969characteristic,inglesfield1983plasmon}. In recent
years, valence band plasmon satellites, which were observed
experimentally in three-dimensional metals and semiconductors
\cite{Guzzo,LischnerFadley,guzzo2014multiple,ley1972x}, but also in
two-dimensional systems, such as doped graphene and semiconductor
quantum-well electron gases
\cite{Rotenberg2,Dial,walter2011effective}, received much
attention.

To analyze and design the properties of plasmon satellites for
photonics and plasmonics applications, an accurate, material-specific
theoretical description of electron-plasmon interactions is
needed. This is achieved by the GW plus cumulant (GW+C) approach
\cite{Langreth,Hedin}, where the cumulant expansion of the electron
Green's function $G$ is truncated at second order in the screened
Coulomb interaction $W$. GW+C calculations yielded good agreement with
experimental photoemission and tunneling spectra in a wide range of
physical systems
\cite{Guzzo,lischner2013physical,lischner2014satellite,LischnerFadley,HedinAryasetiawan,guzzo2014multiple}
and also with highly accurate coupled-cluster Green's function
calculations \cite{mcclain2015spectral}.

While Green's function methods, such as the GW+C approach, often
produce highly accurate results, gaining intuition and insights into
the underlying many-body processes can be difficult. In this paper,
we develop a complementary many-body wavefunction-based approach for
plasmonic (and more generally, bosonic) satellites in the electron
spectral function which offers a clear and simple physical picture of
electron-plasmon interactions and leads to new insights into the
results of GW+C calculations. Specifically, this approach reveals that
the concepts of satellite dispersion, satellite lineshape and
satellite linewidth are closely related, explains why in
three-dimensional materials the plasmon satellite band structure looks
like a shifted copy of the quasiparticle band structure and
demonstrates that previous models of plasmon satellites in three
dimensions are over-simplified and cannot be applied to
lower-dimensional systems. We present results for three-dimensional
[silicon and the three-dimensional electron gas (3DEG)],
two-dimensional (doped graphene) and one-dimensional [the
one-dimensional electron gas (1DEG)] systems.

\emph{Green's function theory}.---The electron spectral function is
related to many observables, such as the tunneling and photoemission
spectrum, and the contribution $A^{\text{IP}}_{\bm{k}}(\omega)$ (with
$\bm{k}$ denoting the wave vector and we omit a band index)
describing the removal of an electron is given by
\cite{MahanBook,giuliani2005quantum}
\begin{align}
  A^{\text{IP}}_{\bm{k}}(\omega)= \sum_\lambda |\langle N-1,\lambda |
  c_{\bm{k}} | \text{GS} \rangle |^2 \delta(\omega+E_{N-1,\lambda}-E_{\text{GS}}),
\label{eq:A}
\end{align}
where $|\text{GS}\rangle$ and $E_{\text{GS}}$ denote the ground state
wave function and energy of the $N$-electron system, respectively, and
$|N-1,\lambda \rangle$ and $E_{N-1,\lambda}$ denote the eigenstates
and energies of the $(N-1)$-electron system.

The spectral function is related to the one-electron Green's function
$G_{\bm{k}}(\omega)$ via $A_{\bm{k}}(\omega)=1/\pi \times
|\text{Im}G_{\bm{k}}(\omega)|$. Within the generalized GW+C approach,
the retarded Green's function is expressed as function of time $t$ via
\cite{kas2014cumulant}
\begin{align}
  G_{\bm{k}}(t) = -i \Theta(t) e^{-i E^{\text{HF}}_{\bm{k}}t + C_{\bm{k}}(t)},
\end{align}
where $E^{\text{HF}}_{\bm{k}}$ denotes the Hartree-Fock orbital energy
(given by
$E^{\text{HF}}_{\bm{k}}=\epsilon_{\bm{k}}+\Sigma^{\text{X}}_{\bm{k}}-V^{xc}_{\bm{k}}$
with $\epsilon_{\bm{k}}$, $V^{xc}_{\bm{k}}$ and
$\Sigma^{\text{X}}_{\bm{k}}$ denoting the mean-field orbital energy,
the mean-field exchange-correlation potential and the bare exchange
self energy, respectively). Also, $C_{\bm{k}}(t)$ is the cumulant
function given by
\begin{align}
  C_{\bm{k}}(t) = \frac{1}{\pi} \int d\omega
  |\text{Im}\Sigma_{\bm{k}}(\omega + E_{\bm{k}})| \frac{e^{-i\omega
      t} + i\omega t - 1}{\omega^2},
\label{eq:C}
\end{align}
where $\Sigma_{\bm{k}}(\omega)$ denotes the $\text{GW}$ self energy
\cite{hedin1967new,HedinBook} and $E_{\bm{k}}$ is the GW quasiparticle
energy.

To gain physical understanding, it is useful separate the cumulant
function into a satellite contribution $C^{\text{sat}}_{\bm{k}}(t)$,
which contains the $e^{-i\omega t}$ term in Eq.~\eqref{eq:C}, and a
quasiparticle contribution, which contains the $(i\omega t -1)$
term. Expanding the Green's function in powers of
$C^{\text{sat}}_{\bm{k}}$ leads to a representation of the spectral
function as the sum of a quasiparticle contribution $A^{\text{qp}}_{\bm{k}}$
and an infinite series of plasmon satellite contributions
$A^{(m)}_{\bm{k}}$ (with $m$ denoting the number of plasmons that
are created in the shake-up process). Specifically, the first satellite
contribution can be expressed as
\begin{align}
  A^{(1)}_{\bm{k}}(\omega) = \int d\omega' C^{\text{sat}}_{\bm{k}}(\omega -
  \omega') A^{\text{qp}}_{\bm{k}}(\omega').
\label{eq:A_1}
\end{align}
Approximating $A^{\text{qp}}_{\bm{k}}(\omega) \approx Z_{\bm{k}}
\delta^{(\Gamma_{\bm{k}})}(\omega - E_{\bm{k}})$ with $Z_{\bm{k}}$
denoting the renormalization factor and $\delta^{(\Gamma)}$ being a
Lorentzian of width $\Gamma$, we find that $ A^{(1)}_{\bm{k}}(\omega)
\approx Z_{\bm{k}}/\pi \times
\text{Im}\Sigma_{\bm{k}}(\omega)/(\omega-E_{\bm{k}})^2$.  

Evaluating Eq.~\eqref{eq:A_1} requires the calculation of the imaginary
part of the GW self energy. To clarify the physical picture, we use
the self energy of a homogeneous electron system in $D$ dimensions
with a plasmon-pole model for the dielectric response.  With these
assumptions, the electron-removal part of the self energy is given by
\cite{HedinBook}
\begin{align}
  \text{Im} \Sigma^{\text{IP}}_{\bm{k}}(\omega) = \frac{\pi}{L^D}\sum_{\bm{q}}
  \lambda_{\bm{q}} v_{\bm{q}} \delta(\omega -
  E_{\bm{k}-\bm{q}}+\omega_{\bm{q}}),
  \label{eq:ImSig}
\end{align} 
where $v_{\bm{q}}$ denotes the Coulomb interaction in $D$ dimensions
and $\omega_{\bm{q}}$ and $\lambda_{\bm{q}}$ are the plasmon
dispersion relation and the plasmon strength, respectively. Also, $L$
is the linear extension of the system and $\bm{k}-\bm{q}$ corresponds
to a hole state.

Inserting Eq.~\eqref{eq:ImSig} into the expression for $A^{(1)}_{\bm{k}}$
yields
\begin{align}
  A^{(1)}_{\bm{k}}(\omega) = \frac{Z_{\bm{k}}}{L^D} \sum_{\bm{q}}
  \frac{g^2_{\bm{q}}}{(E_{\bm{k}}-E_{\bm{k}-\bm{q}}-\omega_{\bm{q}})^2}
    \delta(\omega - E_{\bm{k}-\bm{q}}+\omega_{\bm{q}}),
\label{eq:jdos}
\end{align}
where we introduced the electron-plasmon coupling strength
$g^2_{\bm{q}} = \lambda_{\bm{q}}v_{\bm{q}}$. Eq.~\eqref{eq:jdos}
shows that the satellite contribution to the spectral function closely
related to the coupling-strength weighted electron-plasmon
joint-density of states
$J_{\bm{k}}(\omega)=1/L^D \times \sum_{\bm{q}}g^2_{\bm{q}}\delta(\omega -
E_{\bm{k}-\bm{q}}+\omega_{\bm{q}})$ comprising only plasmon-hole
pairs with total momentum $\bm{k}$.

\emph{Wavefunction theory}.---We will now demonstrate that the
expression for the satellite contribution to the spectral function
from GW+C [Eq.~\eqref{eq:jdos}] can also be
derived by considering the effective electron-plasmon Hamiltonian
\begin{align}
  H_{el-pl} =& \sum_{\bm{k}} E_{\bm{k}} c^\dagger_{\bm{k}}c_{\bm{k}} +
  \sum_{\bm{q}} \omega_{\bm{q}} a^\dagger_{\bm{q}} a_{\bm{q}} \\ &+
  \sum_{\bm{q},\bm{k}} \frac{g_{\bm{q}}}{\sqrt{L^D}} c^\dagger_{\bm{k}-\bm{q}} c_{\bm{k}} (
  a_{\bm{q}} + a^\dagger_{-\bm{q}}),
\end{align}
where $c_{\bm{k}}$ and $a_{\bm{q}}$ are destruction operators for
quasiparticles and plasmons, respectively. In this Hamiltonian, the
first term describes a set of non-interacting quasiparticles, the
second term describes a set of non-interacting plasmons (or more
generally, bosons) and the third term captures the interaction
between quasiparticles and plasmons. 

This electron-boson Hamiltonian plays a fundamental role in the theory
of electron-phonon interactions, but computing accurate spectral
functions is difficult \cite{MahanBook,Lundqvist,HedinReview}. At
intermediate coupling strengths, different types of perturbation
theory give significantly different results: When compared to highly
accurate path-integral calculations, the self-consistent
Brillouin-Wigner perturbation theory yields substantially \emph{worse}
results than standard Rayleigh-Schr{\"o}dinger perturbation theory
\cite{MahanBook}. 

For electron-plasmon interactions, Lundqvist demonstrated
\cite{Lundqvist} that the application of Brillouin-Wigner perturbation
theory to $H_{el-pl}$ results in the Dyson equation of the GW
approach. Solving this equation, he found \emph{two} solutions. While
the first solution corresponds to a standard quasiparticle excitation, he
assigned the second solution to a novel particle, the
\emph{plasmaron}, a strongly coupled, coherent hole-plasmon
state. Despite several reports claiming the observation of the
plasmaron \cite{Dial,Rotenberg2}, it has become clear recently that no
such excitation exists in known materials and that its spurious
prediction signals the inability of the GW method (and, equivalently,
the Brillouin-Wigner perturbation theory) to describe plasmon
satellites
\cite{Guzzo,lischner2013physical,lischner2014satellite,LischnerFadley}.

Motivated by its accurate description of electron-phonon interactions,
we now apply Rayleigh-Schr{\"o}dinger perturbation theory to
$H_{el-pl}$. Without electron-plasmon interactions, i.e. for
$g_{\bm{q}}=0$, the eigenstates of the $(N-1)$-electron system are
simply $c_{\bm{k}}|\text{GS}\rangle$ (with energy
$E_{\text{GS}}-E_{\bm{k}}$) and
$a^\dagger_{\bm{q}}c_{\bm{k}-\bm{q}}|\text{GS}\rangle$ (with energy
$E_{\text{GS}}-E_{\bm{k}-\bm{q}}+\omega_{\bm{q}}$) \footnote{we
  neglect states with more than one plasmon excitation}. Only the
state $c_{\bm{k}}|\text{GS}\rangle$ gives a contribution to
Eq.~\eqref{eq:A} and the resulting spectral function has a single
delta-function peak and no satellite features.

\begin{figure*}[ht]
  \includegraphics[width=18.cm]{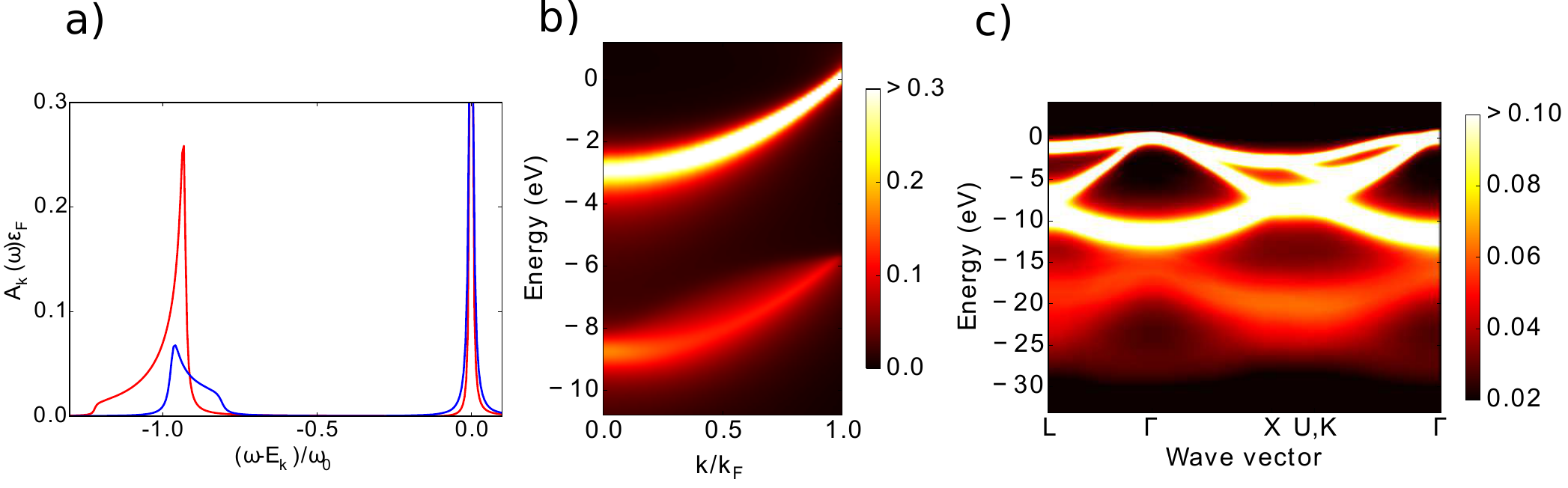}
  \caption{a) GW plus cumulant spectral functions of the
    three-dimensional electron gas at $\bm{k}=0$ for $r_s=1.0$ (red
    curve) and $r_s=3.0$ (blue curve). b) Spectral functions (in 1/eV) of the
    three-dimensional electron gas at $r_s=4.0$ (corresponding to
    metallic sodium) from GW plus cumulant theory. c) Spectral
    functions (in 1/eV) of silicon from \emph{ab initio} GW plus cumulant theory
    calculations.}
  \label{fig:3deg}
\end{figure*}

Next, we include electron-plasmon interactions using first-order
Rayleigh-Schr{\"o}dinger perturbation theory. The non-interacting
states that lie in the energy region of the first satellite are the
hole-plasmon pairs
$a^\dagger_{\bm{q}}c_{\bm{k}-\bm{q}}|\text{GS}\rangle$. Including the
hole-plasmon coupling yields
\begin{align}
  &a^\dagger_{\bm{q}}c_{\bm{k}-\bm{q}}|\text{GS}\rangle \rightarrow
  \nonumber \\ 
 &\left[ a^\dagger_{\bm{q}}c_{\bm{k}-\bm{q}} + \frac{1}{\sqrt{L^D}}
  \frac{g_{\bm{q}}}{E_{\bm{k}}-E_{\bm{k}-\bm{q}}+\omega_{\bm{q}}}c_{\bm{k}}
   + ... \right] | \text{GS} \rangle,
\label{eq:pt}
\end{align}
i.e. the hole-plasmon pair state acquires a quasiparticle component, which
makes this state ``visible'' in the spectral function as a satellite
structure. Inserting Eq.~\eqref{eq:pt} into Eq.~\eqref{eq:A}, we recover
Eq.~\eqref{eq:jdos} for the plasmon satellite contribution to the
electron spectral function. This analysis shows clearly that no single,
coherent hole-plasmon state is formed, but instead the satellite is
comprised of a large number of incoherent, weakly interacting
hole-plasmon pairs. 

\emph{Plasmon satellites in three dimensions.}---We first study the
properties of plasmon satellites in the 3DEG. In this system, the
plasmon dispersion is parabolic at small wave vectors,
i.e. $\omega_{\bm{q}}=\omega_0 + \beta q^2$ \cite{giuliani2005quantum}
and we assume that also the quasiparticle dispersion is parabolic,
i.e. $E_{\bm{k}}=\alpha k^2$.

With these assumptions, we can \emph{analytically} compute the
coupling-constant weighted electron-plasmon joint density of states,
which is closely related to the satellite contribution
$A^{(1)}_{\bm{k}}(\omega)$ to the spectral function \footnote{see
  appendix for details}. For $\bm{k}=0$, we find
\begin{align}
  J_{\bm{k}=0}(\omega) = \frac{\omega_0}{2\pi}
  \frac{ \Theta \left( \text{sgn}(\alpha-\beta)[\omega+\omega_0]
    \right) }{\sqrt{|(\alpha-\beta)(\omega+\omega_0)|}}. 
\label{eq:Asat}
\end{align}
This result shows that the satellite feature is peaked at
$\omega=-\omega_0$, i.e. the satellite is shifted from the
quasiparticle energy by the lowest plasmon energy
$\omega_0$. Moreover, the satellite exhibits a highly
\emph{asymmetric} lineshape, which depends sensitively on the
\emph{relative magnitudes of the plasmon and quasiparticle effective
  masses} [given by $m^*_{pl}=1/(2\beta)$ and $m^*_{qp}=1/(2\alpha)$,
respectively]: if $\beta$ is larger $\alpha$, the satellite peak has a
tail towards higher binding energies (i.e. away from the Fermi
energy). If $\alpha$ is greater than $\beta$, the tail is towards
lower binding energies. If the effective masses are equal, the
satellite structure is symmetric.

Equation~\eqref{eq:Asat} predicts the occurrence of a drastic change
in the satellite lineshape as function of the Wigner-Seitz radius
$r_s$. While the quasiparticle effective mass only has a weak
dependence on $r_s$ and may be approximated by its non-interacting
value, i.e. $\alpha=0.5$ (in atomic units) \cite{HedinBook}, the
plasmon effective mass depends sensitively on $r_s$. Within the
random-phase approximation (RPA), we find $\beta \approx
0.64/\sqrt{r_s}$ \cite{giuliani2005quantum}. At small $r_s$, $\beta$
is large and the tail of the satellite extends to higher binding
energies. For $r_s \gtrsim 1.6$, $\beta$ is smaller than $\alpha$ and
the tail of the satellite extends to lower binding
energies. Figure~\ref{fig:3deg}(a) shows the GW+C spectral functions
for the 3DEG with $r_s=1.0$ and $r_s=3.0$ obtained with a plasmon-pole
model. It can clearly be seen that the tails of the satellites extend
into different directions.

In combination with angle-resolved photoemission spectroscopy (ARPES),
the above analysis imposes useful limits on the value of the plasmon
effective mass. While ARPES experiments do not measure the plasmon
dispersion, they can determine both the quasiparticle effective mass
and the satellite lineshape. Depending on the direction of the
satellite tail [see Fig.~\ref{fig:3deg}(a)], the plasmon effective
mass must be either smaller or larger than the quasiparticle effective
mass. This approach is particularly useful in multiband systems,
where each band leads to an additional constraint on the plasmon
effective mass.

The satellite lineshape from \emph{full} GW+C calculations is more
symmetric than predicted by Eq.~\eqref{eq:Asat} since additional
broadening mechanisms arising from finite quasiparticle linewidths
[see Eq.~\eqref{eq:A_1}] and finite plasmon linewidths (for example,
caused by Landau damping \cite{sturm1982electron}) are taken into
account. The total linewidth of the satellite may thus be approximated
as the sum of the widths of the coupling-strength weighted
hole-plasmon joint density of states, the quasiparticle spectral
function and the plasmon lineshape.  In spectroscopic experiments on
real samples, additional phonon and disorder broadening as well as
broadening due to extrinsic losses occur
\cite{HedinAlmbladh,hedin1998transition}.

Also at nonzero wave vectors, the peak of $J_{\bm{k}}(\omega)$ is
located at an energy $\omega_0$ below the quasiparticle energy
$E_{\bm{k}}$ (see appendix). In other words: the
satellite band is a rigidly shifted copy of the quasiparticle
band. Surprisingly, this means that the \emph{effective mass of the
  satellite is the same as the quasiparticle effective mass}
irrespective of the plasmon effective mass. We confirm our conclusions
by carrying out GW+C calculations of the 3DEG at $r_s=4.0$
(corresponding to metallic sodium) using the frequency-dependent RPA
dielectric function, see Fig.~\ref{fig:3deg}(b).

Generalizing our findings for the plasmon satellite properties of the
3DEG to real materials is straightforward. Fig.~\ref{fig:3deg}(c)
shows the spectral functions of crystalline silicon obtained from
\emph{ab initio} GW+C calculations \footnote{Our calculations employ a
  DFT-LDA mean-field starting point (obtained using the QUANTUM
  ESPRESSO program package\cite{QuantumEspresso}) and a full-frequency
  treatment of the dielectric matrix. Self-energy calculations were
  carried out using the BerkeleyGW program package \cite{BGWpaper}. We
  employ a $8 \times 8 \times 8$ k-point grid, a $5$~Ry dielectric
  cutoff and sampled frequencies up to 300~eV.}. Because of the
parabolic dispersion of the valence quasiparticle bands near the band
extrema and the parabolic dispersion of the plasmon, the plasmon
satellite band structure appears as a rigidly shifted copy of the
quasiparticle band structure, but significantly broadened.

Previous models of plasmon satellites in three-dimensional systems
\cite{Guzzo,caruso2015spectral,caruso2015band} assumed that plasmon
dispersion is a minor effect and approximated the satellite simply as
a shifted, broadened copy of the quasiparticle peak. Such approaches
fail to describe the asymmetric lineshape of the satellite and also
cannot be applied straightforwardly to lower-dimensional systems,
which we discuss below.

\emph{Plasmon satellites in two and one dimensions.}---In
three-dimensional systems, the satellite feature is separated from the
quasiparticle peak by the lowest plasmon frequency. In metallic
\emph{two-dimensional} systems, the plasmon energy is proportional to
the square root of the plasmon wave vector, i.e. $\omega_{\bm{q}}
=\beta\sqrt{q}$ \cite{giuliani2005quantum}, and it is not a priori
clear where the satellite peak is located.

We now apply our GW+C-based analysis of plasmon satellite properties to
\emph{two-dimensional systems} and choose electron-doped graphene as a
test case. Within the Dirac model approach, the two bands in the
vicinity of the Fermi energy are described by a linear dispersion
relation, i.e. $E_{\bm{k}}=\pm v_F k$ with $v_F$ denoting the graphene
Fermi velocity. Here, $\bm{k}$ is measured from the $K$ or $K'$ points
of the graphene Brillouin zone.

Taking into account that electrons in the upper Dirac band give the
dominant contribution to the satellite spectral function at the Dirac
point \cite{polini2008plasmons}, we find that $J_{\bm{k}=0}(\omega)
\propto \Theta(\omega+\tilde{\omega})/\sqrt{\omega + \tilde{\omega}}$,
where $\tilde{\omega}=\beta^2/(4v_F)$ is the separation between the
quasiparticle and satellite peaks. Again, the plasmon satellite
lineshape is highly asymmetric. The dependence of $\beta$ on the
charge density $n$, $\beta \propto \sqrt{n}$
\cite{lischner2014satellite}, gives rise to small changes in the
lineshape as function of the carrier density. Comparing the expression
for $J_{\bm{k}=0}$ of doped graphene to the result for the 3DEG [see
Eq.~\eqref{eq:Asat}], we observe that no drastic changes in the
asymmetry of the lineshape occur as function of the carrier density.

\begin{figure}
  \includegraphics[width=8.cm]{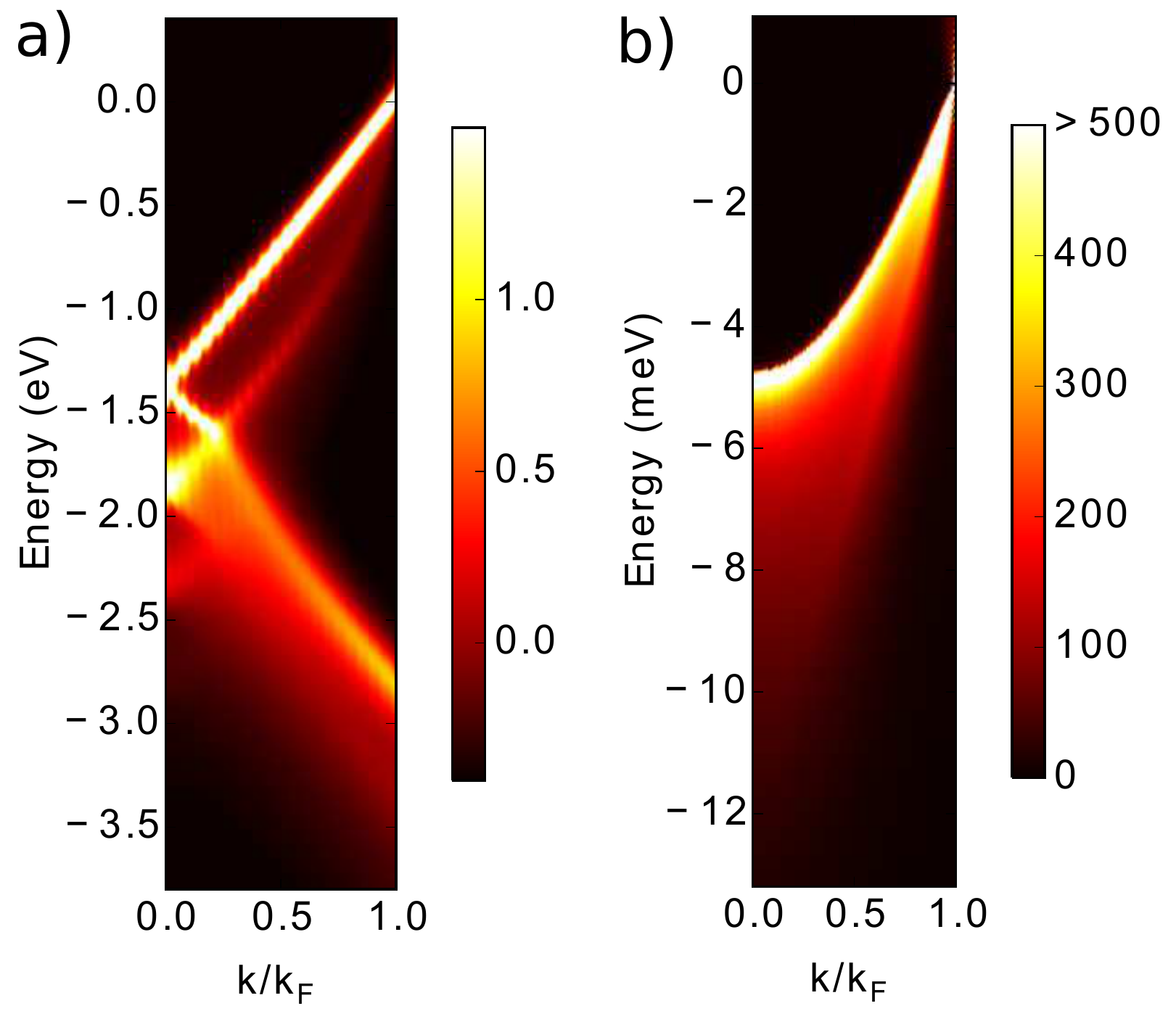}
  \caption{a) Spectral functions (in 1/eV) of doped graphene on a
    silicon carbide substrate from GW plus cumulant theory. b)
    Spectral functions (in 1/eV) of a one-dimensional electron gas
    from GW plus cumulant theory.}
  \label{fig:2d}
\end{figure}

Figure~\ref{fig:2d}(a) shows the spectral functions of doped graphene
on a silicon carbide substrate from GW+C calculations \footnote{Our
  calculations employ a Dirac-model starting point and use the
  frequency-dependent RPA dielectric constant of doped graphene
  \cite{Stauber}. For the dielectric response of the silicon carbide
  substrate, we use the approach of
  Ref. \cite{lischner2013physical}. The Fermi energy is
  $\epsilon_F=1$~eV.}. We observe that the plasmon satellite band is
not a shifted copy of the quasiparticle band, but that the two bands
merge at the Fermi wave vector $k_F$.

Finally, we analyze the plasmon satellite properties in a
\emph{one-dimensional} metallic system, the 1DEG.  In this system, the
plasmon dispersion relation at long wavelengths is given by
$\omega_{\bm{q}}=\beta q \sqrt{\log(1/ql)}$ with $l$ denoting a
cutoff distance \cite{sommerfeld1899fortpflanzung,hu1993many,chaplik1980two}. Assuming a parabolic
quasiparticle dispersion, i.e. $E_{\bm{k}}=\alpha k^2$, we find again
that the first plasmon satellite exhibits a highly asymmetric
lineshape. Fig.~\ref{fig:2d}(b) shows GW+C spectral functions for the
1DEG at $r_s=1.4$ \footnote{We employ a one-dimensional Coulomb
  interaction corresponding to a cylindrical system of radius 100
  $\AA$, an effective mass of $m^*=0.067$ (in atomic units) and a
  background dielectric constant of 12.7 corresponding to a gallium
  arsenide quantum-wire structure \cite{hu1993many}}. In contrast to
graphene and the 3DEG, the plasmon satellite is relatively weak and
appears as a shoulder-like feature near a strong quasiparticle band.

\emph{Summary.}---By connecting the GW+C Green's function approach to
a wavefunction-based perspective, we established the coupling-constant
weighted electron-plasmon joint density of states $J_{\bm{k}}(\omega)$
as a useful quantity for analyzing plasmon satellites in electron
spectral functions. We evaluated $J_{\bm{k}}(\omega)$ for systems in
one, two and three dimensions and demonstrated how the properties of
plasmon satellites are related to the properties of the underlying
electrons and plasmons emphasizing the importance of plasmon
dispersion. Our formalism for electron-plasmon interactions can be
generalized straightforwardly to study the generation of hot
electron-hole pairs in plasmonic devices for photovoltaics and
photocatalysis in the future.

\emph{Acknowledgments.}---The authors would like to thank
Prof. Feliciano Giustino for valuable discussions. J.L. acknowledges
support from EPRSC under Grant No. EP/N005244/1 and also from the
Thomas Young Centre under Grant No. TYC-101. Via J.L.'s membership of
the UK's HEC Materials Chemistry Consortium, which is funded by EPSRC
(EP/L000202), this work used the ARCHER UK National Supercomputing
Service. S.G.L. acknowledges support from the National Science
Foundation under grant DMR15-1508412.

\appendix
\section*{Appendix}

\emph{Electron-Plasmon Joint Density of States in Three
  Dimensions.}---We calculate the coupling-strength weighted joint
density of states $J_{\bm{k}}(\omega)$ comprising only plasmon-hole
pairs with total momentum $\bm{k}$ for a three-dimensional homogeneous
electron gas (3DEG). As shown in the main text, this quantity is
closely related to the first satellite contribution to the electron
spectral function. Specifically, $J_{\bm{k}}(\omega)$ is given by
\begin{align}
  J_{\bm{k}}(\omega) = \int \frac{d^3q}{(2\pi)^3} g^2_{\bm{q}}
  \delta(\omega - E_{\bm{k}-\bm{q}}+\omega_{\bm{q}}),
\end{align}
where $g_{\bm{q}}$ denotes the electron-plasmon coupling strength,
$\omega_{\bm{q}}=\omega_0 + \beta q^2$ is the plasmon dispersion and
$E_{\bm{k}}=\alpha k^2$ is the quasiparticle dispersion. 

Using a plasmon-pole model that conserves sum rules, we find
$g^2_{\bm{q}}=v_{\bm{q}} \omega_0^2/(2\omega_{\bm{q}}) \approx
v_{\bm{q}} \omega_0/2$ with $v_{\bm{q}}=4\pi/q^2$.

For the special case of $\bm{k}=0$, we find
\begin{align}
  J_{\bm{k}=0}(\omega) &= \frac{\omega_0}{\pi} \int_0^\infty dq \delta(\omega +
  \omega_0 - [\alpha-\beta]q^2) \\ \nonumber
  & = \frac{\omega_0}{2\pi} \frac{\Theta( \text{sgn}(\alpha-\beta)(\omega+\omega_0))}{\sqrt{|(\alpha-\beta)(\omega+\omega_0)|}},
\end{align}
which has a peak at $-\omega_0$.

In the general case of nonzero $\bm{k}$, we have to evaluate
\begin{align}
 & J_{\bm{k}}(\omega) = \\ \nonumber
 &\frac{\omega_0}{2\pi} \int_0^\infty dq
  \int_{-1}^1 du \delta(\omega + \omega_0 - \alpha k^2 + 2\alpha ku + [\beta-\alpha]q^2).
\end{align}

Using that $\int_{-1}^1 du \delta(A+Bu)=\Theta(|B|-|A|)/|B|$, we find
that
\begin{align}
 & J_{\bm{k}}(\omega) = \\ \nonumber
 &\frac{\omega_0}{4\pi|\alpha| k} \int_0^\infty
  \frac{dq}{q} \Theta(2kq|\alpha| - |\omega + \omega_0 -\alpha k^2 + [\beta-\alpha]q^2|).
\end{align}

We now assume that $\beta-\alpha > 0$ and distinguish the two cases:
i) $\omega^*_k \equiv \omega + \omega_0 - \alpha k^2 > 0$ and ii)
$\omega^*_k < 0$. To find the position of the peak of
$J_{\bm{k}}(\omega)$, it is sufficient to consider case i) and we find
that
\begin{align}
  J_{\bm{k}}(\omega) = &\frac{\omega_0}{4\pi|\alpha| k} \int_0^\infty
  \frac{dq}{q} \Theta(2kq|\alpha| - \omega^*_k - [\beta-\alpha]q^2) \\
  \nonumber
  & = \frac{\omega_0}{4\pi|\alpha| k}\Theta(1-f_k) \log\left[ \frac{1
    + \sqrt{1-f_k}}{1-\sqrt{1-f_k}}\right],
\label{eq:A1}
\end{align}
with $f_k = [\beta-\alpha]\omega^*_k/(k|\alpha|)^2$. This function
diverges as $\omega^*_k \rightarrow 0$ indicating that
$J_{\bm{k}}(\omega)$ is peaked at $-[\omega_0 - \alpha k^2]$.

For case ii), we have to evaluate
\begin{align}
  J_{\bm{k}}(\omega) &= \frac{\omega_0}{4\pi |\alpha| k} \left[
    \int_0^{q^*} \frac{dq}{q}  \Theta(2kq|\alpha| +
    \omega^*_k + [\beta-\alpha]q^2) + \right. \\ \nonumber
     & \left. \int_{q^*}^\infty   \frac{dq}{q}\Theta(2kq|\alpha| 
    -\omega^*_k - [\beta-\alpha]q^2)\right] \\ \nonumber
  &=\frac{\omega_0}{4\pi |\alpha| k} \log\left[
    \frac{1+\sqrt{1-f_k}}{-1+\sqrt{1-f_k}}\right],
\label{eq:A2}
\end{align} 
with $q^* = \sqrt{|\omega^*_k|/[\beta-\alpha]}$.

Note that the solutions above also describe the case of $\beta-\alpha
<0$, but now Eq.~(5) describes negative $\omega^*_k$ and Eq.~(6)
describes positive $\omega^*_k$. 

\bibliography{paper}
\end{document}